%% file: arimizel_v4.tex
\documentclass[twocolumn,prl]{revtex4-2}%

\usepackage{graphicx}
\usepackage{epstopdf}

\usepackage{amsmath}

\usepackage{times}
\begin{document}

\title{Spatial quantum error correction threshold} 
\date{\today}
\author{Ari Mizel}
\affiliation{Laboratory for Physical Sciences, College Park, Maryland 20740, USA}
\email{ari@arimizel.com}
\begin{abstract}
We consider a spatial analogue of the quantum error correction threshold.  Given individual time-independent subsystems in which quantum information is coherent over sufficiently long lengths, we show how the information can be kept coherent for arbitrarily long lengths by forming time-independent composite systems. The subsystem coherence length exhibits threshold behavior.  When it exceeds a length $\xi_{th}$, meaningful information can be extracted from the ground state of the composite system.  Otherwise, the information is garbled.  The threshold transition implies that the parent Hamiltonian of the ground state has gone from gapped to gapless.  Ramifications of the construction for PEPS and for adiabatic quantum computation are noted.
\end{abstract}
\maketitle

The quantum error correction \cite{Shor95,Gaitan08,Lidar13} threshold theorem \cite{Aharonov08,Knill98,Kitaev97} is a foundational element of the theory of quantum computation.  It states that, provided a system has quantum gates with infidelity below a threshold value $p_{th}$, scalable computation is possible.  The noise in the gates can be tamed using redundancy, by encoding physical qubits into logical qubits.  Roughly speaking, if quantum information in individual subsystems remains coherent for sufficiently long times, it can be kept coherent for arbitrarily long times merely by forming composite systems.

In this paper, we consider the spatial analogue of this phenomenon.  Given individual subsystems in which quantum information is coherent over sufficiently long lengths, can it be kept coherent for arbitrarily long lengths merely by forming composite systems?  Is there spatial threshold behavior?  In defining the question, it is important to emphasize that we are considering the properties of a {\em time-independent} quantum state.  In the usual, temporal version of quantum error correction, it is often supposed that the qubits occupy different spatial locations.  A set of time-dependent errors can therefore be visualized as occurring at a set of distinct positions, a perspective that is especially helpful in the context of topological codes \cite{Dennis02}.  However, this is only superficially similar to the question we are asking about time-independent spatial coherence.

To answer the question, we specify a quantum subsystem that serves as an elemental building block, analogous to a qubit in the usual quantum error correction context.  Given any quantum circuit $\mathbf{c}$, it is possible to encode a fault-tolerant version of $\mathbf{c}$ into the time-independent state $|\Psi(\theta)\rangle$ of an assembly of these subsystems.  A parameter $\theta$ tunes the minimum coherence length $\xi(\theta)$ of the subsystems.  We show that $\xi(\theta)$ exhibits spatial quantum error correction threshold behavior.  When $\xi(\theta)$ is just over a threshold value $\xi_{th}$, the output of $\mathbf{c}$ can be extracted from $|\Psi(\theta)\rangle$.  Otherwise, the output of $\mathbf{c}$ is generally too garbled to extract.  Conveniently, $|\Psi(\theta)\rangle$ is the ground state of a 2-local parent Hamiltonian $H(\theta)$.  The construction leverages ground-state quantum computation \cite{Mizel01,Mizel02,Mizel04,Mizel21} with important new features from quantum error correction.

The paper is structured as follows.  The bulk of the exposition spells out the construction of $|\Psi(\theta)\rangle$ and $H(\theta)$.  We derive the threshold behavior as the coherence length $\xi(\theta)$ crosses just over $\xi_{th}$.  Then, we show that there is a gapped to gapless transition in $H(\theta)$.   The conclusion discusses some implications.

To describe the construction of $|\Psi(\theta)\rangle$, we assume that our starting quantum circuit $\mathbf{c}$ is composed of only initializations and unitary gates.  We can encode physical qubits into logical qubits to form a fault-tolerant \cite{Aharonov08,Knill98,Kitaev97} quantum circuit $\mathbf{C}$.  While measurements are often used within fault-tolerant circuits to extract entropy, this is inessential  \cite{Aharonov08}; it will be convenient to assume that our circuit $\mathbf{C}$ uses only initializations and one- and two-qubit unitary gates.  The fault-tolerance allows $\mathbf{C}$ to produce the correct output of $\mathbf{c}$ even if each of the gates of $\mathbf{C}$ is replaced by a perfect gate followed with probability $p$ by a depolarizing channel.  It is only necessary that $p \le p_{th} - \delta p $, where $p_{th}$ is the quantum error correction threshold and $\delta p$ is fixed and positive. 

\begin{figure}[htb]
\begin{tabular}{cc}
\begin{tabular}{l} (a)\\ \includegraphics[width=0.225\textwidth]{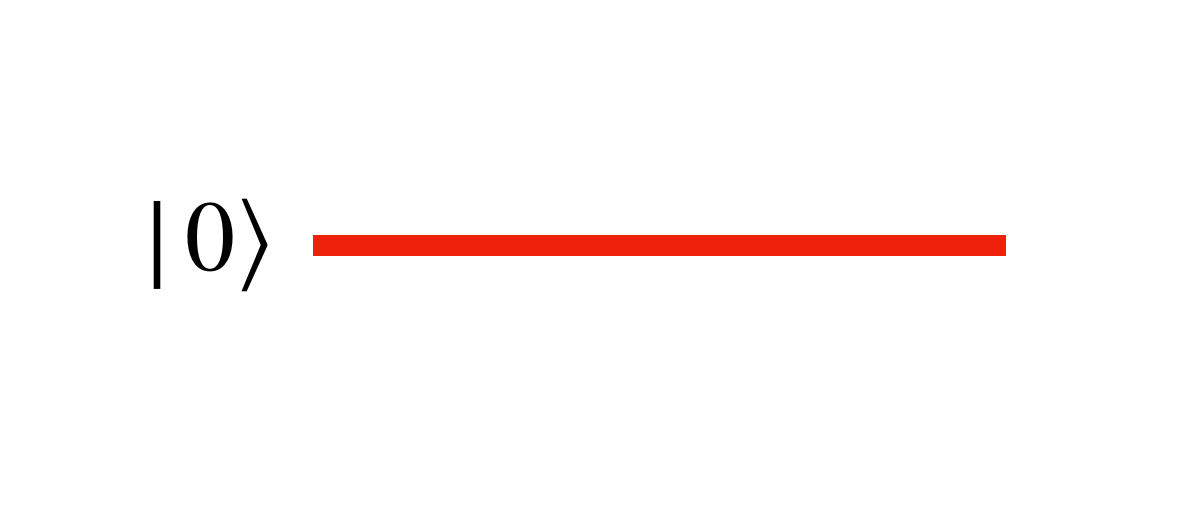} \end{tabular} & \begin{tabular}{l}  (c) \\ \includegraphics[width=0.225\textwidth]{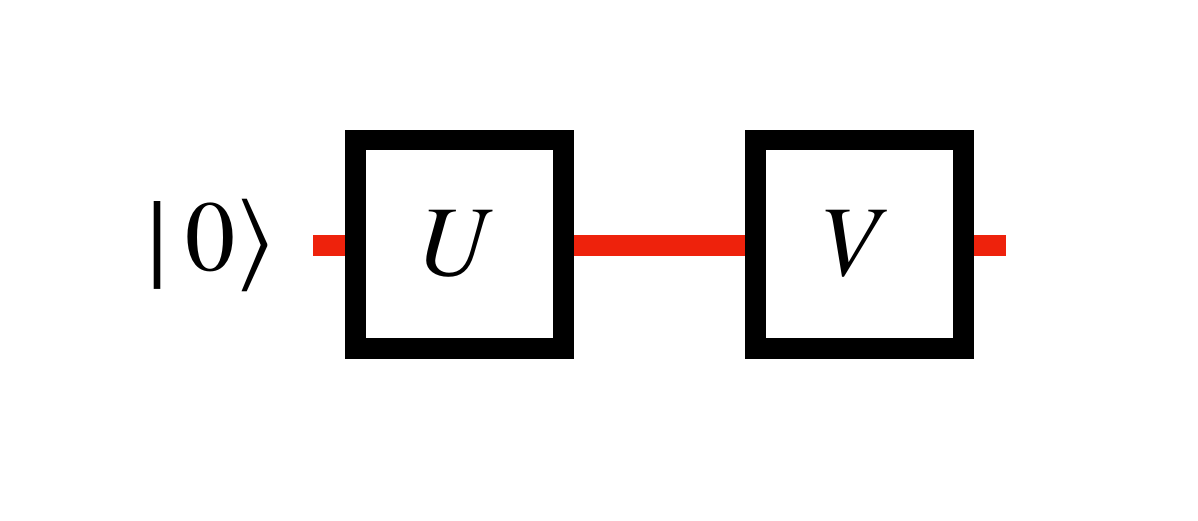}   \end{tabular} \\
 \begin{tabular}{l} (b) \\ \includegraphics[width=0.225\textwidth]{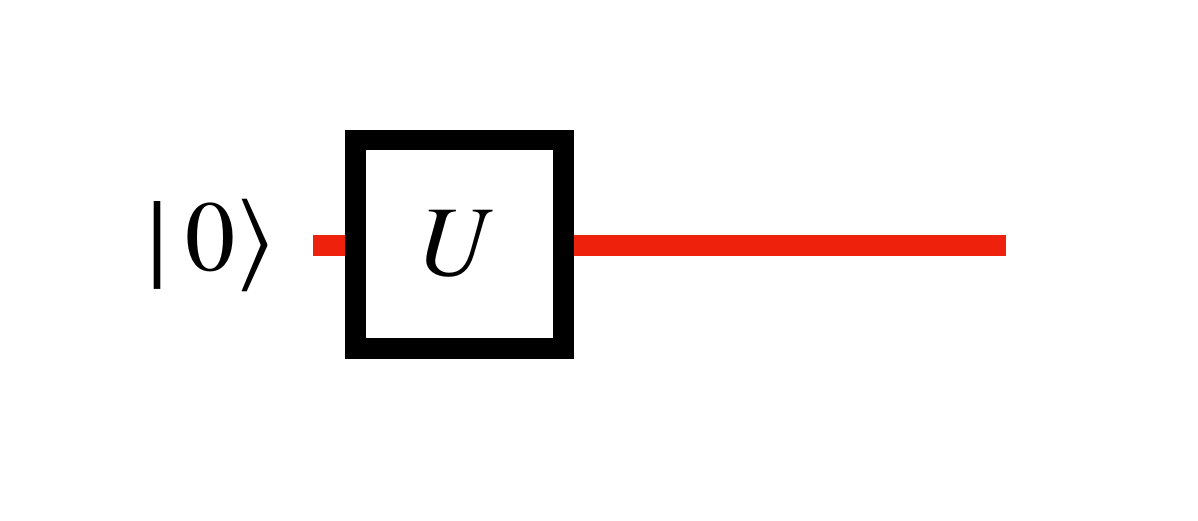} \end{tabular} & \begin{tabular}{l} (d)\\\includegraphics[width=0.225\textwidth]{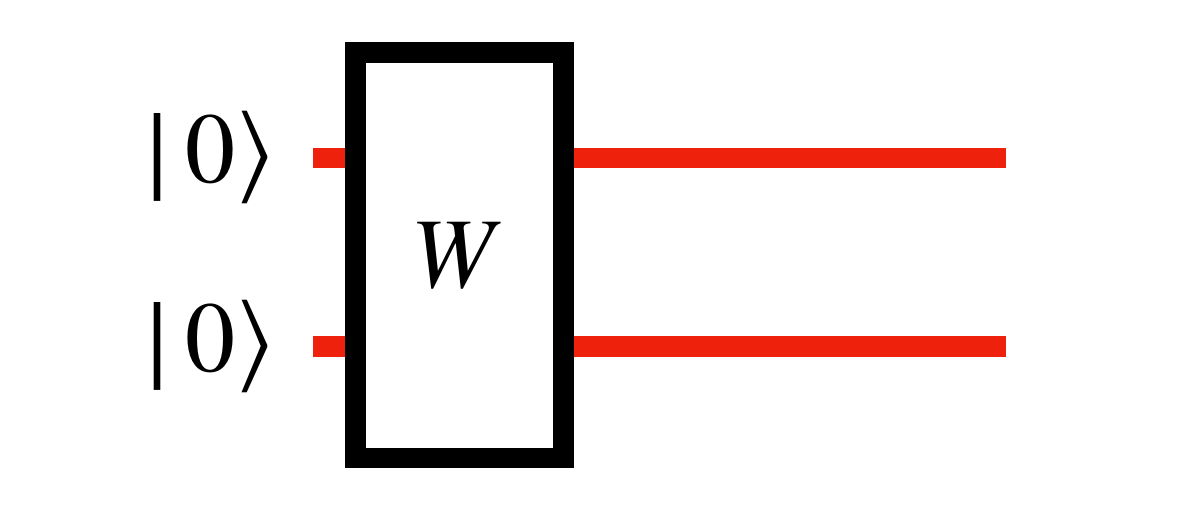} \end{tabular} \end{tabular}
   
  \caption{Example circuits, with time flowing to the right.  (a) Trivial circuit composed solely of initialization without any gates.  (b) Initialization of qubit, followed by a one-qubit gate $U$.  (c) Circuit (b) followed by a second one-qubit gate, $V$.  (d) Initialization of 2 qubits followed by two-qubit gate.}
\label{fig:circuits}
\end{figure}

\begin{figure}[htb]
\begin{tabular}{ccc}
 \begin{tabular}{l} (a)\\ \includegraphics[width=0.125\textwidth]{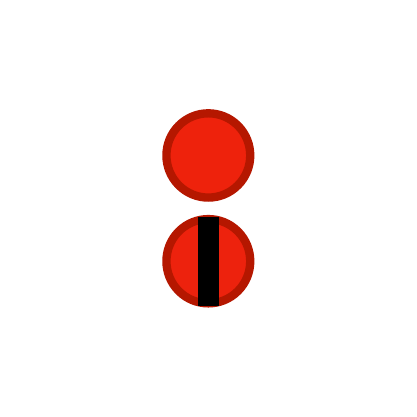}  \\ (b) \\ \includegraphics[width=0.125\textwidth]{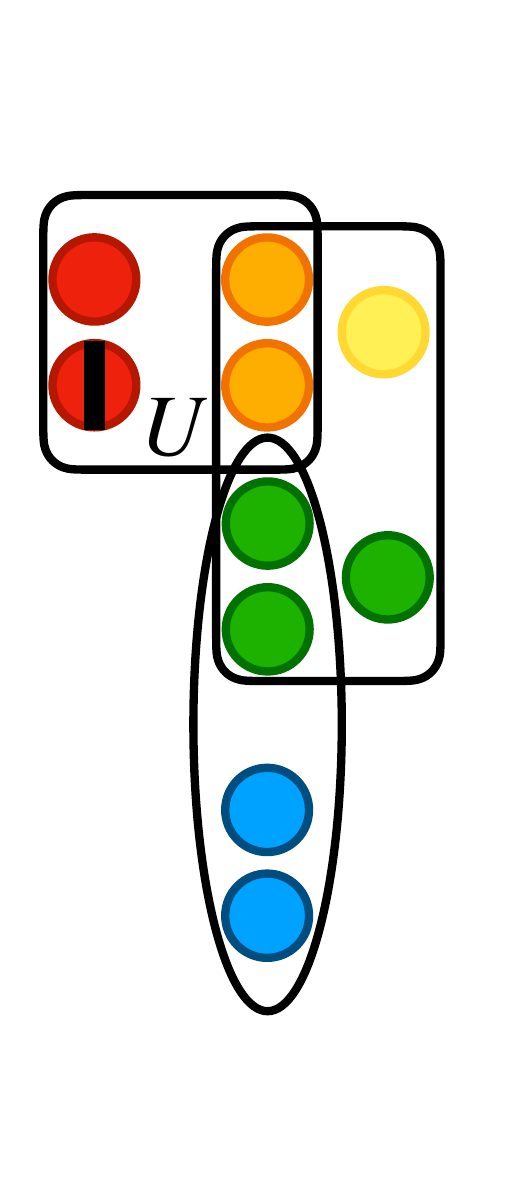}   \end{tabular} &\hspace{-0.125in}
 \begin{tabular}{l} (c) \\ \includegraphics[width=0.15\textwidth]{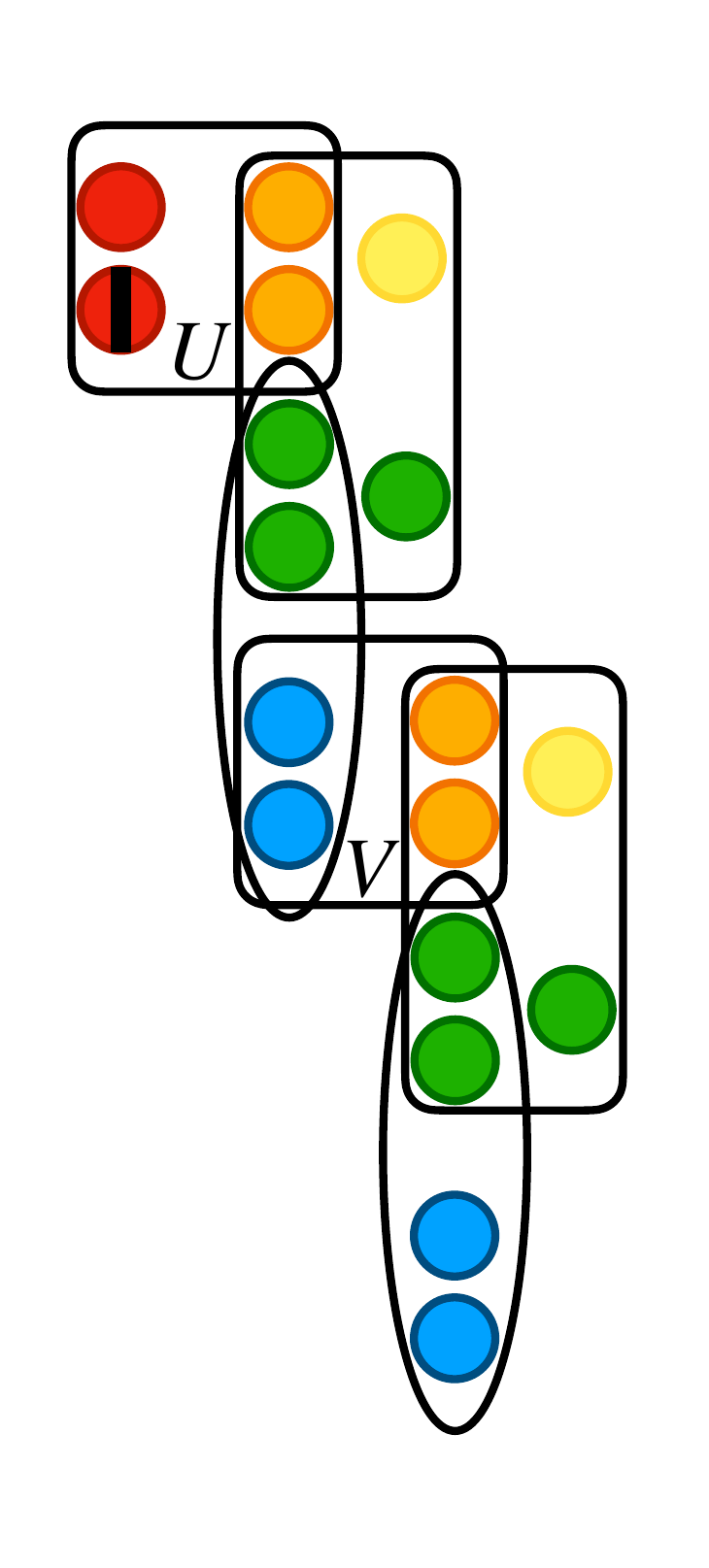} \end{tabular} & \begin{tabular}{l} (d)\\\includegraphics[width=0.125\textwidth]{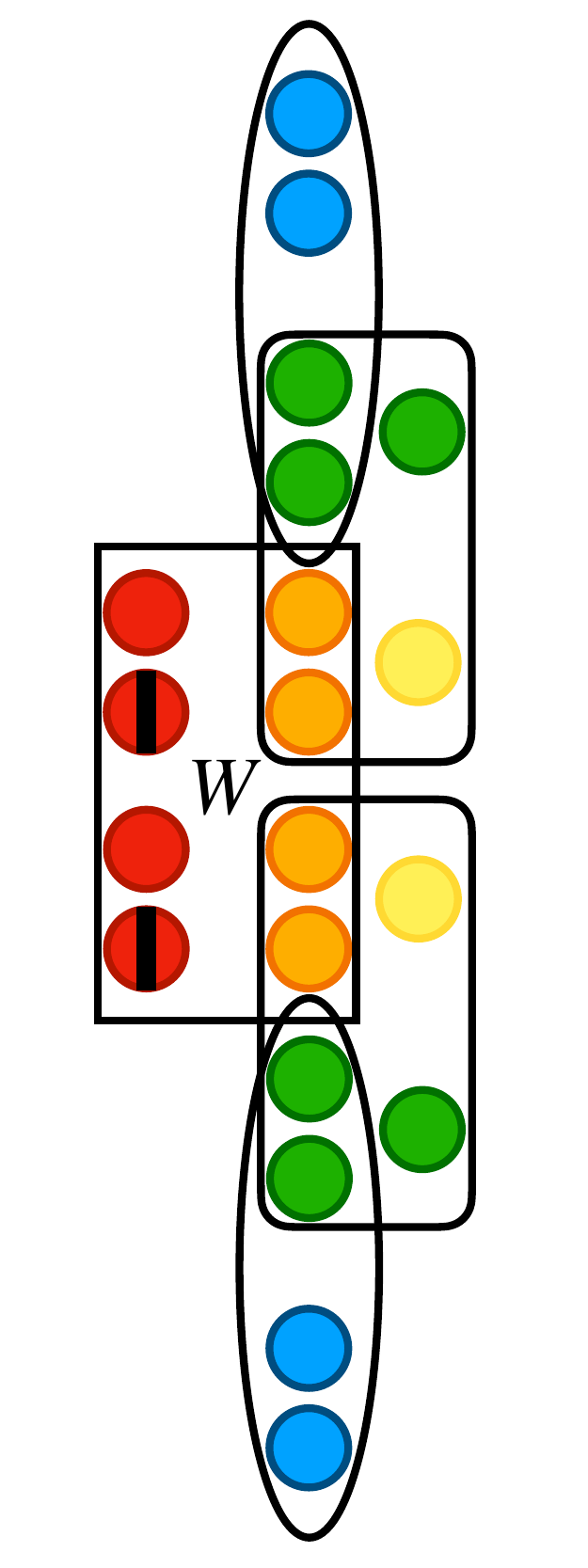} \end{tabular} 
 \end{tabular}
  \caption{Graphical depiction of Hamiltonians corresponding to circuits in Fig. \ref{fig:circuits}, with space flowing to the right rather than time.   (a) A pair of circles represents the $2$ states $\{|0_0\rangle,|1_0\rangle\}$.   Concretely, one can think of an electron shared between quantum dots or a Cooper pair shared between superconducting islands.  Vertical black line represents the energy penalty $H_{init}.$  (b) To apply a gate $U$, the $2$ dimensional Hilbert space is extended to $2 \otimes 3 \otimes (1\oplus 2\oplus 2)$ dimensions.  Colored circles represent the associated basis states according to the correspondance blue $\otimes$ green $\otimes ($yellow $\oplus$ orange $\oplus$ red$)$.  The terms of $H^U_{one}(\theta)$ are depicted using a square outline to represent Eq. (\ref{eq:HU}), a rectangle for (\ref{eq:HP}), and an oval for (\ref{eq:HB}).  (c) To apply a second gate $V$, the leftmost part of the $2 \otimes 3 \otimes (1\oplus 2\oplus 2)$ dimensional space is extended, yielding a $2 \otimes 3 \otimes (1\oplus 2\oplus 2) \otimes 3 \otimes (1\oplus 2\oplus 2)$ dimensional space. (d) Analogue of (b) for a two-qubit gate $W$.}
  \label{fig:Hamiltonians}
\end{figure}

The map from $\mathbf{C}$ to  $|\Psi(\theta)\rangle$ is most easily described using explicit circuit examples as shown in Fig. \ref{fig:circuits}.   For each example, we will also specify a 2-local parent Hamiltonian $H(\theta)$ as
a sum of initialization terms $H_{init}$, one-qubit gate terms $H_{one}^{U}(\theta)$, and two-qubit gate terms $H_{two}^{W}(\theta)$ in one-to-one correspondence with the initializations and gates of $\mathbf{C}$.  The Hamiltonian will be represented symbolically in Fig. \ref{fig:Hamiltonians}.

Start with the trivial circuit in Fig. \ref{fig:circuits}(a) that simply initializes a qubit.   Define a $2$ dimensional Hilbert space with basis $\left\{\left|0_0\right>,\left|1_0\right>\right\}$ where the ket $\left| b_s\right>$ has ``bit" value $b$ and computational ``stage" value $s$.  For the trivial circuit of Fig. \ref{fig:circuits}(a), there is only stage $s=0$, and the desired time-independent state is $\left|0_0\right>$.  It is the zero-energy ground state of the positive semi-definite parent Hamiltonian $H_{init} = \epsilon \left|1_0\right>\left<1_0\right|$ with $\epsilon$ a fixed energy scale.  Fig. \ref{fig:Hamiltonians}(a) depicts $H_{init}$ symbolically.

To apply a one-qubit gate $U$ to the qubit after initialization, as in Fig. \ref{fig:circuits}(b), extend its 2 dimensional Hilbert space so that it has dimension $2 \otimes 3 \otimes (1 \oplus 2 \oplus 2)$.  
 We will explain this extension in 2 steps.  Consider first extending from a 2 dimensional space to a $2 \oplus 2$ dimensional space, replacing the original basis $\left\{\left|0_0\right>,\left|1_0\right>\right\}$ with an extended basis $\left\{\left|0_1\right>,\left|1_1\right>\right\} \cup \left\{\left|0_0\right>,\left|1_0\right>\right\}$. The time-independent state of the qubit in this extended space is assigned the  form ${ \frac{1}{\sqrt{2}}(\left|0_1\right>  \left<0\right| U \left|0\right>  +  \left|1_1\right>\left<1\right| U \left|0\right>+\left|0_0\right> )}$, where $ \left<0\right| U \left|0\right>$ and $\left<1\right| U \left|0\right>$ are matrix elements of the one-qubit gate $U$.
This state of the qubit is a superposition of computational stage $s=0$  after initialization and stage $s=1$ after $U$ is applied.
We define the operator ${\mathcal U} = \sum_{b,\beta,s = 0,1}\left|b_s\right> \left<b \right| U \left| \beta \right> \left<{\beta}_s\right|$ that applies $U$ while keeping the stage fixed; then our state can be written in the compact form $\frac{1}{\sqrt{2}} ({\mathcal U} \left|0_1\right>+\left|0_0\right>)$.
 
\begin{figure}[htb]
\includegraphics[width=0.475\textwidth]{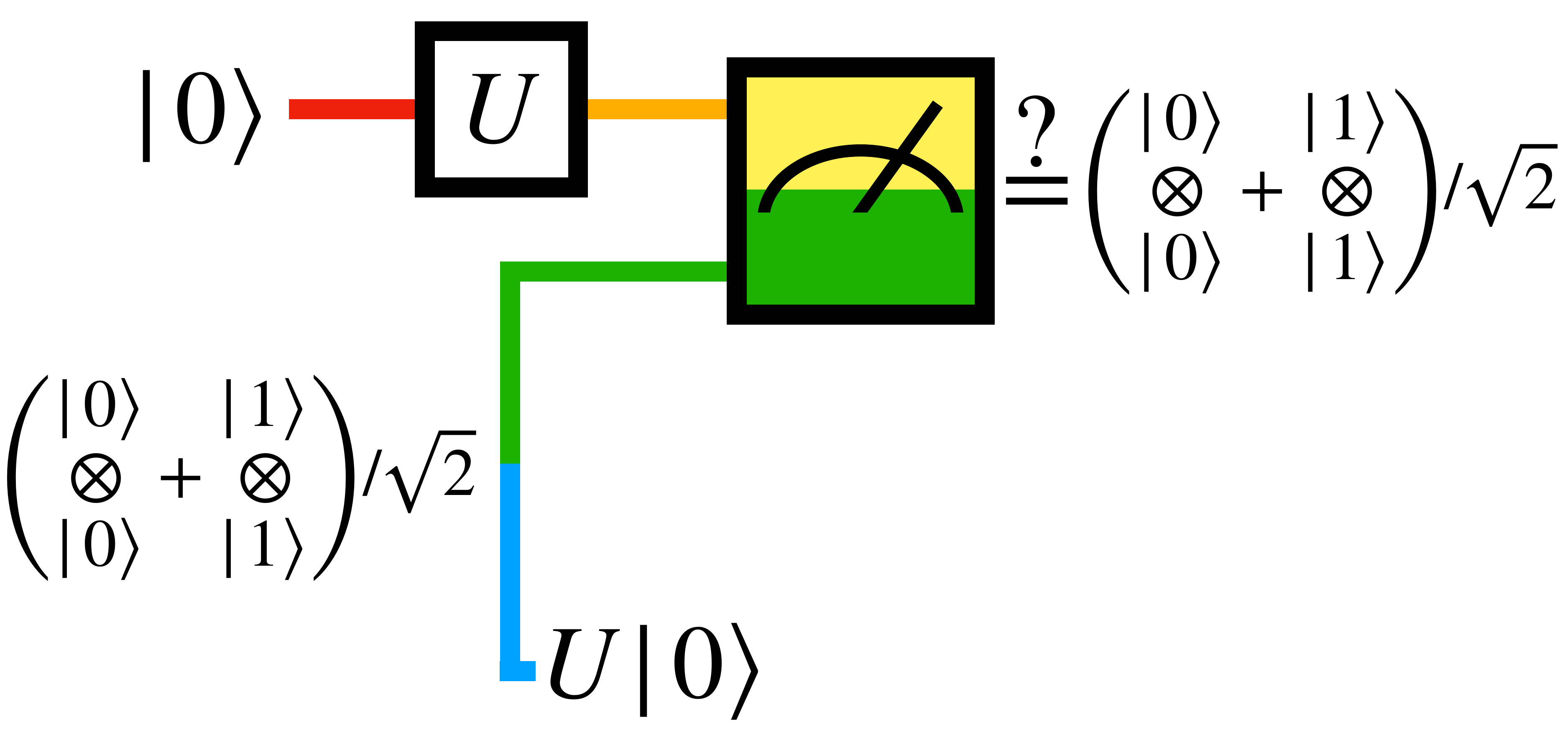} 
  \caption{Teleportation circuit.  A post-selection step checks that the measured state is $(\left|0\right>\otimes \left|0\right>+\left|1\right>\otimes \left|1\right>) /\sqrt{2}$; when this is true, the outgoing state on the bottom is $U|0\rangle$.   The color coding details the correspondence between the parts of the circuit and the parts of Fig. \ref{fig:Hamiltonians}(b).}
  \label{fig:teleportation}
\end{figure}

Next, extend the space of the qubit from ${2 \oplus 2}$ dimensional to  $2 \otimes 3 \otimes (1 \oplus 2 \oplus 2)$ dimensional.  This is done to incorporate a teleportation-like step \cite{Bennett93} acting after $U$.  The teleportation circuit in Fig. \ref{fig:teleportation} is color coded to clarify the role of each part of the extended Hilbert space.  A convenient basis is $\left\{\left|0_0\right>,\left|1_0\right>\right\}\otimes \left\{\left|\mbox{IDLE},\left|0_0\right>,\left|1_0\right>\right>\right\}\otimes \left(\left\{\left|\mbox{IDLE}\right>\right\}\cup\left\{\left|0_1\right>,\left|1_1\right>\right\}\cup\left\{\left|0_0\right>,\left|1_0\right>\right\}\right) $.    Our state is assigned the form $ \left| \psi^{U}(0)\right>$ 
  where
\begin{align}
& \left| \psi^{U}(b)\right> = \sqrt{2/(8 \cos^2\theta + \sin^2\theta)} \,\,\, \times \nonumber \\
&  \Big[ \cos \theta  (\left|0_0\right>\otimes \left|0_0\right>+\left|1_0\right>\otimes \left|1_0\right>) \otimes ( {\mathcal U}\left|b_1\right>+\left|b_0\right>)\nonumber\\ 
&\hspace{0.0in}  +\sin \theta \, {\mathcal U}\left|b_0\right> \otimes \left|\mbox{IDLE}\right>\otimes\left|\mbox{IDLE}\right>/\sqrt{2}  \Big] \label{eq:psi(b)} 
\end{align}
for $b=0,1$.
In the first term of $\left| \psi^{U}(b)\right>$, we prepend, alongside the state $({\mathcal U}\left|b_1\right>+\left|b_0\right>)/\sqrt{2}$,  a Bell-pair $ (\left|0_0\right>\otimes \left|0_0\right>+\left|1_0\right>\otimes \left|1_0\right>)$ needed for teleportation.  The second term $ {\mathcal U}\left|b_0\right> \otimes \left|\mbox{IDLE}\right>\otimes\left|\mbox{IDLE}\right> $ completes teleportation, consuming the original qubit state and half of the Bell-pair, so that the quantum information teleports to the other half of the Bell pair.  When the parameter $\theta$ is close to $\pi/2$, the fidelity of the teleportation is high.

Now that the state $\left| \psi^{U}(0)\right>$ of the qubit inhabits a $2 \otimes 3 \otimes (1\oplus 2 \oplus 2)$ dimensional space, in what sense do we still have a qubit? If we compute the density matrix of the system and trace out the $3 \otimes (1\oplus 2 \oplus 2)$ dimensional part, the leftmost 2 dimensional qubit-sized space contains its quantum information.  To demonstrate this, define the one-qubit ``gate operator'' $\hat{g}_{one}^{U} = \left| \psi^{U}(0)\right>\left<0_0\right|+\left| \psi^{U}(1)\right>\left<1_0\right|$ in terms of (\ref{eq:psi(b)}).  This operator is a mapping from a 2 dimensional space to a $2 \otimes 3 \otimes (1\oplus 2 \oplus 2)$ dimensional space.  Our qubit state is simply $\left| \psi^{U}(0)\right>= \hat{g}_{one}^{U} \left|0_0\right>$.  Define the superoperator $g_{one}^{U}\left(\rho\right) = \mbox{Tr}_{3 \otimes (1 \oplus 2 \oplus 2)}  \hat{g}_{one}^{U} \rho \hat{g}_{one} ^{U \dagger}$.  One calculates that $g_{one}^{U}\left(\rho\right) = (1-p_{one}) \mathcal{U} \rho \mathcal{U}^\dagger + p_{one}\mbox{Tr} \rho \,\,\, I/2$, so that it applies $U$ followed by a depolarizing channel with probability $p_{one}(\theta)= 8 \cos^2 \theta/(8 \cos^2\theta + \sin^2\theta)$.   The reduced density matrix of the leftmost 2 dimensional part of the $2 \otimes 3 \otimes (1 \oplus 2 \oplus 2)$ dimensional space of the qubit is $g_{one}^{U}\left( \left|0_0\right>\left<0_0\right| \right)$.  This yields the desired output $U|0\rangle$ of Fig. \ref{fig:circuits}(b) as $\theta$ approaches $\pi/2$.

The $2 \otimes 3 \otimes (1\oplus 2 \oplus 2)$ dimensional subsystem is the elemental building block of our construction.  If we assume that the quantum state is distributed spatially like the dots laid out in Fig. \ref{fig:Hamiltonians}(b), it is natural to define a coherence length describing the decay of quantum coherence from the rightmost 2 dimensional part of the $2 \otimes 3 \otimes (1\oplus 2 \oplus 2)$ dimensional Hilbert space to the leftmost 2 dimensional part.  In the case of a one-qubit gate like we have been considering, we denote the coherence length $\xi_{one}(\theta)$.  In terms of the trace distance, let $e^{-1/\xi_{one}(\theta)} = {\min_{\rho} 1 - \frac{1}{2}||{\mathcal U} \rho {\mathcal U}^\dagger - g_{one}^{U}\left(\rho\right)||_{tr}} = {1 -\frac{1}{2} p_{one}(\theta)}$.

The positive semi-definite Hamiltonian 
\[
H_{one}^{U}(\theta) +I ^{(2)} \otimes I^{(3)} \otimes H_{init}
\]
has $ \hat{g}_{one}^{U} \left|0_0\right>$ as a non-degenerate zero-energy ground state.   Here, $I^{(k)}$ denotes the identity operator on a $k$-dimensional space, and $H_{one}^{U} (\theta) =  I ^{(2)} \otimes I^{(3)} \otimes H^{{\mathcal U}}+  H_{B} \otimes (I^{(1)}\oplus I^{(2)}\oplus I^{(2)}) + I^{(2)} \otimes H_{P}(\theta)$ with 
\begin{equation}
H^{{\mathcal U}}= \epsilon \sum_{b} (  {\mathcal U} \left|b_1\right> -\left| b_0 \right> ) ( \left<b_1\right| {{\mathcal U}}^\dagger - \left<b_0\right|  )/2 \label{eq:HU},
\end{equation}
enforcing the action of the specific unitary $U$ and
\begin{align}
& H_{B}  = \nonumber \\
& \frac{\epsilon}{2} \left[(\left|1_0\right>\left|0_0\right>-\left|0_0\right>\left|1_0\right>)(\left<1_0\right|\left<0_0\right|-\left<0_0\right|\left<1_0\right|) \right. \nonumber \\
& + (\left|1_0\right>\left|0_0\right>+\left|0_0\right>\left|1_0\right>)(\left<1_0\right|\left<0_0\right|+\left<0_0\right|\left<1_0\right|) \nonumber \\
& \left. + (\left|0_0\right>\left|0_0\right>-\left|1_0\right>\left|1_0\right>)(\left<0_0\right|\left<0_0\right|-\left<1_0\right|\left<1_0\right|) \right] \label{eq:HB}
\end{align}
imposing an energy penalty if the Bell pair in the first term of (\ref{eq:psi(b)}) is not of the desired form $(\left|0_0\right>\otimes \left|0_0\right> + \left|1_0\right>\otimes \left|1_0\right>)/\sqrt{2}$.  Finally,
\begin{align}
& H_{P}(\theta) = \epsilon \nonumber \\
&  \Big[\left( \sin \theta \frac{\left|0_0\right>\left|0_1\right>+\left|1_0\right>\left|1_1\right>}{\sqrt{2}} - \cos \theta \left|\mbox{IDLE}\right>\left|\mbox{IDLE}\right> \right) \nonumber \\
& \hspace{0.05in} \left( \sin \theta \frac{\left<0_0\right|\left<0_1\right|+\left<1_0\right|\left<1_1\right|}{\sqrt{2}}- \cos \theta \left<\mbox{IDLE}\right|\left<\mbox{IDLE}\right|\right) \nonumber\\
& + \left|\mbox{IDLE}\right>\left<\mbox{IDLE}\right| \otimes \sum_{b,s = 0,1} \left|b_s\right>\left<b_s\right| \nonumber \\
& + \sum_{b = 0,1} \left|b_0\right>\left<b_0\right| \otimes \left|\mbox{IDLE}\right>\left<\mbox{IDLE}\right|  \Big]\label{eq:HP} 
\end{align}
effects a projection that mimics the Bell-basis measurement step of teleportation and imposes an energy penalty unless both targets of the measurement undergo the step in tandem.  Fig. \ref{fig:Hamiltonians}(b) sketches the Hamiltonian, emphasizing the domain of each term.

The construction of Fig. \ref{fig:Hamiltonians}(b) can be iterated.
For instance, if a second unitary gate $V$ is applied to our qubit, as in Fig. \ref{fig:circuits}(c), the ground state is assigned the form $(\hat{g}_{one}^{V} \otimes I^{(3)} \otimes (I^{(1)}\oplus I^{(2)}\oplus I^{(2)})) \left| \psi^{U}(0)\right> = (\hat{g}_{one}^{V} \otimes I^{(3)} \otimes (I^{(1)}\oplus I^{(2)}\oplus I^{(2)})) \hat{g}_{one}^{U}|0_0\rangle$.  The action of $\hat{g}_{one}^{V}$ iteratively expands the leftmost 2 dimensional part of the Hilbert space, so that, instead of a $2 \otimes 3 \otimes (1\oplus 2 \oplus 2)$ dimensional space, the qubit now inhabits a $2 \otimes 3 \otimes (1\oplus 2 \oplus 2) \otimes 3 \otimes (1\oplus 2 \oplus 2)$ dimensional space.  The reduced density matrix of the leftmost $2$ dimensional Hilbert space is $g_{one}^{V}(g_{one}^{U}(|0_0\rangle\langle 0_0|))$.  This equals the output produced by a noisy quantum circuit that starts with $|0\rangle$, applies $U$ followed by a depolarizing channel with probability $p_{one}(\theta)$, then applies $V$ followed by a depolarizing channel with probability $p_{one}(\theta)$.  The Hamiltonian, depicted symbolically in Fig. \ref{fig:Hamiltonians}(c), is 
\begin{align*}
& H_{one}^{V}(\theta) \otimes I^{(3)} \otimes  (I^{(1)}\oplus I^{(2)}\oplus I^{(2)}) \\
&+I ^{(2)} \otimes I^{(3)} \otimes \\
& \hspace{0.5in} (H_{one}^{U}(\theta)+ (I^{(1)}\oplus I^{(2)}\oplus I^{(2)}) \otimes I^{(3)} \otimes H_{init}).
\end{align*}

To incorporate the effect of a two-qubit gate $W$ in $\mathbf{C}$, as in Fig. \ref{fig:circuits}(d), $\hat{g}_{one}^{U}$ is replaced with an operator $\hat{g}_{two}^{W}$.  (See Supp. Mat.)  Its associated superoperator $g_{two}^{W}$
applies $W$ followed by a depolarizing channel on one or both qubits with probability  $p_{two}(\theta) = (32 \cos^4\theta + 8 \cos^2 \theta \sin^2\theta)/(32 \cos^4\theta + 8 \cos^2 \theta \sin^2\theta + \sin^4 \theta)$.  The Hamiltonian $H_{two}^{W}(\theta)$ is represented in Fig. \ref{fig:Hamiltonians}(d).  The coherence length $\xi_{two}(\theta)$ associated  with the two qubit gate is defined as $e^{-1/\xi_{two}(\theta)} = {\min_{\rho} 1 - \frac{1}{2}||{\mathcal W} \rho {\mathcal W}^\dagger - g_{two}^{W}\left(\rho\right)||_{tr}} =1 - \frac{3}{4}p_{two}(\theta)$. 

By iterating the constructions above, employing $2 \otimes 3 \otimes (1\oplus 2 \oplus 2)$-dimensional subsystems for each of the gates in $\mathbf{C}$, one obtains a $|\Psi(\theta)\rangle$ that contains the output of $\mathbf{C}$ and its parent Hamiltonian $H(\theta)$.   The state has the form $|\Psi(\theta)\rangle = \dots \hat{g}_{two}^{W}\dots \hat{g}_{one}^{U}\dots |0_0\rangle^{\otimes Q}$ where there is an operator of the form $\hat{g}_{two}^{W}$ for each two-qubit gate in $\mathbf{C}$, an operator of the form $\hat{g}_{one}^{U}$ for each one-qubit gate in $\mathbf{C}$, and $Q$ is the number of qubits in $\mathbf{C}$.  We have omitted tensor products with identity operators, abbreviating, for example, $(\hat{g}_{one}^{V} \otimes I^{(3)} \otimes (I^{(1)}\oplus I^{(2)}\oplus I^{(2)}))$ as $\hat{g}_{one}^{V}$.  The final reduced density matrix of dimension $2^{\otimes Q}$, obtained by tracing out all but the final 2 dimensional Hilbert space of each qubit, takes the abbreviated form $\dots \left(g_{two}^{W}\left(\dots g_{one}^{U}\left( \dots \left(\left|0_0\right>\left<0_0\right|^{\otimes Q} \right) \dots\right)\dots\right)\right)$, where again we have omitted tensor products with identity operators.  This equals the density matrix that would be produced by executing the quantum circuit $\mathbf{C}$ with each perfect unitary followed by depolarization with probability $p_{one}(\theta)$ or $p_{two}(\theta)$.  Because $p_{one}(\theta) \le p_{two}(\theta)$,  we set the gate error probability $p$ to $p_{two}(\theta)$.  Then, $\mathbf{C}$'s fault-tolerance implies the output of $\mathbf{c}$ can be extracted, by decoding the final density matrix of dimension $2^{\otimes Q}$ of $\left|\Psi(\theta)\right>$, provided $p_{two}(\theta) \le  p_{th} - \delta p$.

This gives rise to our spatial quantum error correction threshold.   We set the minimum coherence length $\xi(\theta)$ to $\xi_{two}(\theta)$ because $\xi_{two}(\theta)\le \xi_{one}(\theta)$.  When  
\begin{equation}
\xi_{two}(\theta) \ge \xi_{th} + \delta \xi,
\end{equation}
the output of $\mathbf{c}$ can be extracted.  Here, $\xi_{th}$ is defined as $\xi_{th} = \xi_{two}(\theta_{th})$ where $p_{two}(\theta_{th}) = p_{th}$, and $\delta \xi = \xi_{two}(p^{-1}_{two}(p_{th}+\delta p))-\xi_{th}$.

When $\xi(\theta)$ crosses the threshold, there are ramifications for the properties of $H(\theta)$.  A key property of interest is the energy gap, which is computed by taking the limit of large system size.  Since the form of $H(\theta)$ is determined by a fault-tolerant circuit $\mathbf{C}$, this thermodynamic limit should be taken by specifying a family of larger and larger fault-tolerant circuits.  One natural family comprises circuits associated with a given quantum algorithm of growing problem size.  However, it is simpler to consider \footnote{This example was pointed out to me by M. B. Hastings.} a circuit $\mathbf{c}$ that starts with two qubits initialized to $|0\rangle \otimes |0\rangle$, applies a Hadamard gate to the second qubit to produce $|0\rangle\otimes (|0\rangle+|1\rangle)/\sqrt{2}$ , and then applies a controlled-NOT targeting the first qubit in order to produce the Bell pair $(|0\rangle \otimes |0\rangle + |1\rangle \otimes |1\rangle)/\sqrt{2}$.  Finally,  a string of $G$ identity gates is applied to each qubit of the pair.   To take the limit of large system size, let $G$ grow, with the fault-tolerant circuit $\mathbf{C}$ requiring ever bulkier logical qubits.

\begin{figure}[htb]
\ \includegraphics[width=0.5\textwidth]{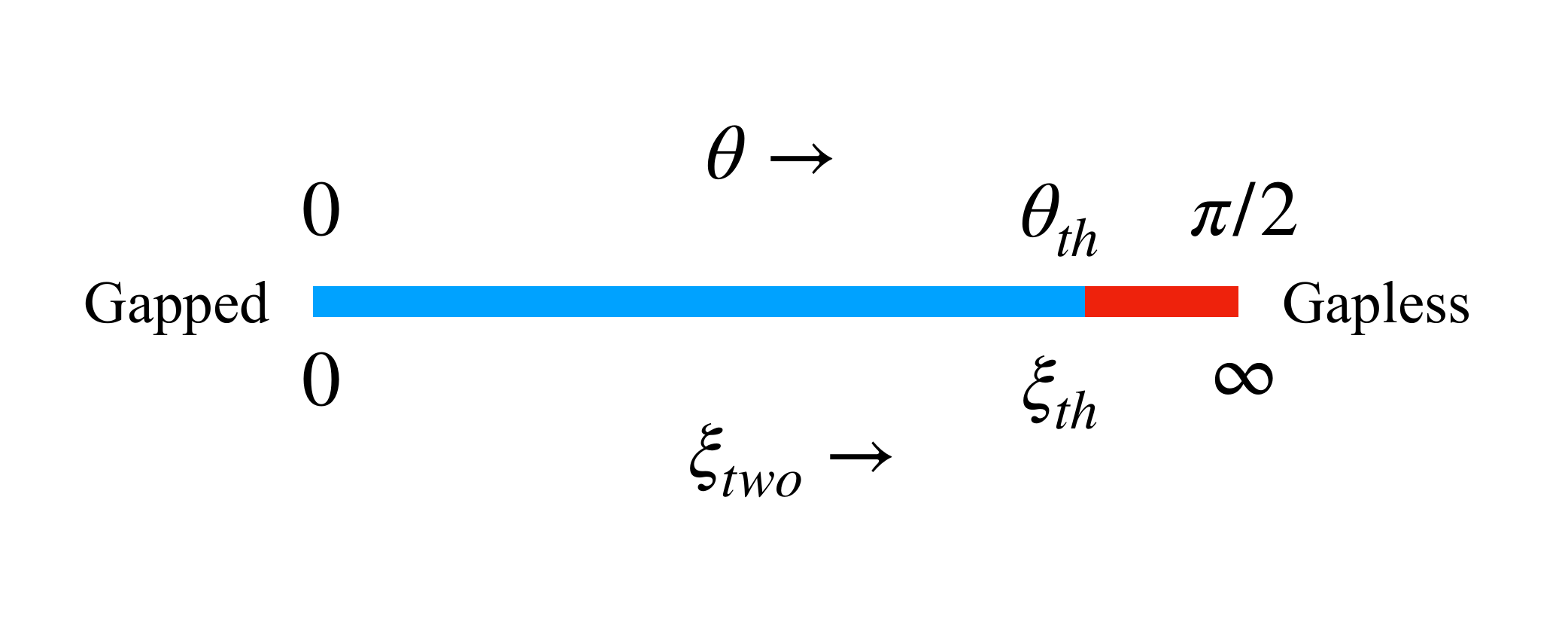} 
\caption{Phase diagram of $H(\theta))$.  Values of $\theta$ appear above the line with the corresponding value of $\xi_{two}$ below.  As $\xi_{two}$ increases from 0 to $\infty$, it crosses the threshold value $\xi_{th}$, and the ground state develops long-range entanglement.   This implies the Hamiltonian has become gapless.}
  \label{fig:phasediagram}
\end{figure}

At the point $\theta = 0$, where $\xi(\theta)=0$, the energy eigenvalues of $H(\theta)$ are easily obtained by inspecting the forms of $H_{one}^{U}(\theta)$ and $H_{two}^{W}(\theta)$.  We find a gap for all $G$.  Now, $H(\theta)$ is going to remain gapped for small values of $\theta$, but, by the time $\xi(\theta)$ crosses $\xi_{th} + \delta \xi$, the system must have undergone a transition to a gapless phase (see Fig. \ref{fig:phasediagram}).  This must happen because $\mathbf{C}$ successfully outputs an intact logical Bell pair for all values of $G$.  Thus, the ground state $|\Psi(\theta)\rangle$ contains long range entanglement between the members of this Bell pair, which can only happen if $H(\theta)$ is gapless \cite{Hastings06,Nachtergaele06}.

This gapped to gapless transition is of independent interest in the context of projected entangled pair states (PEPS) \cite{Cirac21}.  The state $|\Psi(\theta)\rangle$ can be written as a PEPS (see Supp. Mat.).  While 1-dimensional PEPS, referred to as matrix product states, generally have gapped parent Hamiltonians \cite{Fannes92}, understanding when higher-dimensional PEPS parent Hamiltonians are gapped and when they are gapless is a subtle problem under active consideration \cite{Kastoryano19}.  Our $|\Psi(\theta)\rangle$ and $H(\theta)$ provide a useful example to inform this investigation.  

In conclusion, we have demonstrated a spatial quantum error correction threshold.  It is realized in the time-independent ground state $|\Psi(\theta)\rangle$ of a parent Hamiltonian $H(\theta)$.  The threshold behavior is associated with a gapped to gapless transition in $H(\theta)$.  The construction presented in this paper could be considered for use in universal adiabatic quantum computing \cite{Albash2018}.  It enjoys fault-tolerance against implementation errors of the Hamiltonian and leakage errors.  However, the gapless property of $H(\theta)$ indicates that, in thermal equilibrium, the system is not fault-tolerant against excitations.

We gratefully acknowledge helpful comments by M. B. Hastings, M. Kruger, D. A. Lidar, K. Miller, V. Molino, K. D. Osborn, V. N. Smelyanskiy, and M. M. Wilde.

\bibliography{ari_mizelbib2}

\setcounter{figure}{0} \renewcommand{\thefigure}{S\arabic{figure}} 

\setcounter{equation}{0} \renewcommand{\theequation}{S\arabic{equation}} 

\include{arimizel_appendix_v3}

\end{document}

%% file: arimizel_appendix_v3.tex
\onecolumngrid
\section*{Supplemental Material}

\section*{Two-qubit gates}

This section completes the discussion of Fig. \ref{fig:Hamiltonians}(d), detailing the case of a two-qubit gate $W$.  The two participating qubits inhabit a $[2 \otimes 3 \otimes (1\oplus 2 \oplus 2)]\otimes [2 \otimes 3 \otimes (1\oplus 2 \oplus 2)]$ dimensional Hilbert space.  We define 4 states of the system by

\begin{align}
\left|\psi_0^{W}(b,B)\right> =  & \frac{1}{\sqrt{32 \cos^4\theta + 8 \cos^2 \theta \sin^2\theta + \sin^4 \theta}}   \nonumber \\
& \Big[4 \cos^2 \theta \frac{1}{\sqrt{2}}(\left|0_0\right>\left|0_0\right>+\left|1_0\right>\left|1_0\right>)    \left| b_0\right>    \frac{1}{\sqrt{2}} (\left|0_0\right>\left|0_0\right>+\left|1_0\right>\left|1_0\right>)      \left| B_0\right>   \nonumber \\
& +\sum_{b^\prime,B^\prime} \langle b^\prime|\langle B^\prime| W|b\rangle| B\rangle  \Big( 4 \cos^2 \theta \frac{1}{\sqrt{2}}(\left|0_0\right>\left|0_0\right>+\left|1_0\right>\left|1_0\right>)    \left| b^\prime_1\right>    \frac{1}{\sqrt{2}} (\left|0_0\right>\left|0_0\right>+\left|1_0\right>\left|1_0\right>)      \left| B^\prime_1\right> \nonumber \\
&\hspace{1.5in}+2 \cos \theta \sin \theta   \left|b^\prime_0\right> \left|\mbox{IDLE}\right>\left|\mbox{IDLE}\right>\frac{1}{\sqrt{2}} (\left|0_0\right>\left|0_0\right>+\left|1_0\right>\left| 1_0\right>)  \left| B^\prime_1\right> \nonumber \\
&\hspace{1.5in}+2 \cos \theta \sin \theta \frac{1}{\sqrt{2}}(\left|0_0\right>\left|0_0\right>+\left|1_0\right>\left|1_0\right>)    \left| b^\prime_1\right>  \left| B^\prime_0\right>  \left|\mbox{IDLE}\right>\left|\mbox{IDLE}\right>\nonumber \\
&\hspace{1.5in}+ \sin^2 \theta   \left|b^\prime_0\right> \left|\mbox{IDLE}\right>\left|\mbox{IDLE}\right> \left| B^\prime_0\right>\left|\mbox{IDLE}\right>\left|\mbox{IDLE}\right> \Big) \Big]. \label{eq:psi0W}
\end{align}
In this equation, line 1 contains the normalization constant.  Line 2 corresponds to the stage of the computation before $W$ is applied.  The input qubit states $|b\rangle$ and $|B\rangle$ are each accompanied by a Bell pair $(\left|0_0\right>\left|0_0\right>+\left|1_0\right>\left|1_0\right>)/\sqrt{2}$ that will be needed for teleportation.  At line 3, $W$ has been applied, but teleportation has not yet occurred, so the Bell pairs are still present.  At lines 4 and 5, teleportation has occurred for one of the two qubits but not the other.  At line 6, teleportation has occurred for both qubits, completing the gate.

   The gate operator is defined by $\hat{g}_{two}^{W} = \sum_{b,B} \left|\psi_0^{W}(b,B)\right>\langle b_0|\langle B_0|$.  The corresponding superoperator is
\begin{align}
 g_{two}^{W} \left(\rho\right)  & =  \mbox{Tr}_{3 \otimes (1 \oplus 2 \oplus 2)}   \mbox{Tr}_{3 \otimes (1 \oplus 2 \oplus 2)}   \hat{g}_{two}^{W} \rho \hat{g}_{two} ^{W \dagger} \nonumber \\
& = (1-p_{two}(\theta)) \mathcal{W} \rho \mathcal{W}^\dagger \nonumber \\
& + \frac{4 \cos^2 \theta\sin^2 \theta}{32 \cos^4\theta + 8 \cos^2 \theta \sin^2\theta + \sin^4 \theta}  \sum_{b,b^\prime,B}\langle b|\langle B|\rho|b^\prime\rangle|B\rangle   \mathcal{W} \Big( |b_0\rangle  \langle b_0^\prime| \otimes  \frac{|0_0\rangle\langle 0_0|+|1_0\rangle\langle 1_0| }{2}  \Big)  \mathcal{W}^\dagger \nonumber \\
& + \frac{4 \cos^2 \theta\sin^2 \theta}{32 \cos^4\theta + 8 \cos^2 \theta \sin^2\theta + \sin^4 \theta} \sum_{b,B,B^\prime}\langle b| \langle B|\rho|b\rangle |B^\prime\rangle  \mathcal{W} \Big( \frac{|0_0\rangle\langle 0_0|+|1_0\rangle\langle 1_0| }{2}  \otimes    |B_0\rangle  \langle B_0^\prime| \Big) \mathcal{W}^\dagger \nonumber \\
& + \frac{32 \cos^4 \theta}{32 \cos^4\theta + 8 \cos^2 \theta \sin^2\theta + \sin^4 \theta} \mbox{Tr} \rho \frac{|0_0\rangle\langle 0_0|+|1_0\rangle\langle 1_0| }{2} \otimes  \frac{|0_0\rangle\langle 0_0|+|1_0\rangle\langle 1_0| }{2} \label{eq:gtwoW}  
\end{align}
with $p_{two}(\theta) = (32 \cos^4\theta + 8 \cos^2 \theta \sin^2\theta)/(32 \cos^4\theta + 8 \cos^2 \theta \sin^2\theta + \sin^4 \theta)$.   Here, ${\mathcal W} = \sum_{s,b,B,b^\prime,B^\prime} |b_s\rangle|B_s\rangle \langle b| \langle B| W|b^\prime \rangle |B^\prime\rangle\langle b^\prime_s|\langle B^\prime_s|$ applies $W$ while keeping the stage variable fixed.  The second line of Eq. (\ref{eq:gtwoW}) corresponds to the successful application of the gate $W$.   In the third and fourth lines, a depolarization channel has been applied to just one of the two qubits.  In the final line, depolarization channels have been applied to both qubits.   In Fig. \ref{fig:Hamiltonians}(d), the reduced density matrix of the two qubits, after tracing out the  $3 \otimes (1 \oplus 2 \oplus 2)$ dimensional part of the Hilbert space of each, is $g_{two}^{W}\left( \left|0_0\right>|0_0\rangle \langle 0_0| \left<0_0\right| \right)$.  This yields the desired output $W|0\rangle|0\rangle$ of Fig. \ref{fig:circuits}(d) as $\theta$ approaches $\pi/2$.

The parent Hamiltonian of the ground states (\ref{eq:psi0W}) has the form 
\begin{align}
H_{two}^{W}(\theta) = &\frac{\epsilon}{2} \sum_{b,B} I^{(2)} \otimes I^{(3)} \otimes | b_0\rangle  \langle b_0| ] \otimes   [I^{(2)} \otimes I^{(3)}  \otimes |B_0\rangle \langle B_0| ] \\
& +\frac{\epsilon}{2}\sum_{b,B} [ I^{(2)} \otimes I^{(3)} \otimes | b_1\rangle  \langle b_1| ] \otimes   [I^{(2)} \otimes I^{(3)}  \otimes |B_1\rangle \langle B_1| ]  \nonumber\\
& - \frac{\epsilon}{2}\sum_{b,B,b^\prime,B^\prime}\langle b^\prime| \langle B^\prime| W | b\rangle |B\rangle [ I^{(2)} \otimes I^{(3)} \otimes | b^\prime_1\rangle \langle b_0|] \otimes  [ I^{(2)} \otimes I^{(3)} \otimes  |B^\prime_1\rangle \langle B_0|]   \nonumber\\
& - \frac{\epsilon}{2}\sum_{b,B,b^\prime,B^\prime}\langle b| \langle B| W^\dagger | b^\prime \rangle |B^\prime\rangle [ I^{(2)} \otimes I^{(3)} \otimes | b_0\rangle \langle b^\prime_1|] \otimes  [ I^{(2)} \otimes I^{(3)} \otimes  |B_0\rangle \langle B^\prime_1|]   \nonumber\\
& + \frac{\epsilon}{2}   [ I^{(2)} \otimes I^{(3)} \otimes \sum_{b} | b_0\rangle \langle b_0|] \otimes  [ I^{(2)} \otimes I^{(3)} \otimes (|\mbox{IDLE}\rangle \langle \mbox{IDLE}|+\sum_{B} | B_1\rangle \langle B_1|)] \nonumber \\
& + \frac{\epsilon}{2}    [ I^{(2)} \otimes I^{(3)} \otimes (|\mbox{IDLE}\rangle \langle \mbox{IDLE}|+\sum_{b} | b_1\rangle \langle b_1|)] \otimes [ I^{(2)} \otimes I^{(3)} \otimes \sum_{B} | B_0\rangle \langle B_0|]  \nonumber \\
& +  [H_{B} \otimes (I^{(1)}\oplus I^{(2)}\oplus I^{(2)}) + I^{(2)} \otimes H_{P}] \otimes [I^{(2)} \otimes I^{(3)} \otimes (I^{(1)}\oplus I^{(2)}\oplus I^{(2)})] \nonumber\\
& +  [I^{(2)} \otimes I^{(3)} \otimes (I^{(1)}\oplus I^{(2)}\oplus I^{(2)})]  \otimes [H_{B} \otimes (I^{(1)}\oplus I^{(2)}\oplus I^{(2)}) + I^{(2)} \otimes H_{P}]. \nonumber
\end{align}
The first four lines are analogous to the single-qubit gate case (\ref{eq:HU}), despite superficial complexity resulting from the tensor product notation.  Both qubits move together from stage 0 to stage 1, undergoing the gate $W$.  The next two lines impose an energy penalty if either qubit attempts to traverse the gate alone.  The seventh line is concerned with the teleportation of one qubit, and the final line is concerned with the teleportation of the other qubit.  These last two lines employ the Hamiltonians (\ref{eq:HB}) and (\ref{eq:HP}).

The circuit in Fig. \ref{fig:circuits}(d) includes initializations.  Thus, the total Hamiltonian for Fig. \ref{fig:Hamiltonians}(d) is 
\begin{align}
H(\theta) = & H_{two}^{W}(\theta) +[I^{(2)} \otimes I^{(3)} \otimes (I^{(1)}\oplus I^{(2)}\oplus I^{(2)})] \otimes [I^{(2)} \otimes I^{(3)} \otimes H_{init}] \nonumber \\
 &+ [I^{(2)} \otimes I^{(3)} \otimes H_{init}] \otimes [I^{(2)} \otimes I^{(3)} \otimes (I^{(1)}\oplus I^{(2)}\oplus I^{(2)})].
 \end{align}

\section*{PEPS form of ground state}

Here, we show how the ground state $|\Psi(\theta)\rangle$ can be written as a projected entangled pair state (PEPS).  Rather than a formal proof, which would require the introduction of cumbersome notation, we consider the examples shown in Fig. \ref{fig:Hamiltonians}.  The general pattern will become clear from these examples. 

In the case of Fig. \ref{fig:Hamiltonians}(a), the ground state is a trivial PEPS: $|\Psi(\theta)\rangle = |0_0\rangle$.  For Fig. \ref{fig:Hamiltonians}(b), 
define the map
\begin{align}
\hat{A}  = & \cos \theta (|0_0\rangle \langle 0_0|+|1_0\rangle \langle 1_0|) \otimes   (|0_1\rangle \langle 0_1|+|1_1\rangle \langle 1_1|)  \nonumber \\
 & +\frac{1}{\sqrt{2}} \sin \theta  (|\mbox{IDLE}\rangle\langle 0_0| \otimes  |\mbox{IDLE}\rangle\langle 0_1|+ |\mbox{IDLE}\rangle\langle 1_0|\otimes |\mbox{IDLE}\rangle\langle 1_1|).
\end{align}
Then, the ground state $|\Psi(\theta)\rangle$ has the PEPS form  $(I^{(2)} \otimes \hat{A} )[(|0_0\rangle |0_0\rangle + |1_0\rangle |1_0\rangle ) \otimes  ({\mathcal U}|0_1\rangle+|0_0\rangle)]$ up to normalization.  In Fig. \ref{fig:Hamiltonians}(c),  $|\Psi(\theta)\rangle$ has the PEPS form  
\begin{align}
(I^{(2)} \otimes \hat{A} \otimes \hat{A})[(|0_0\rangle |0_0\rangle + |1_0\rangle |1_0\rangle ) \otimes (({\mathcal V}|0_1\rangle+|0_0\rangle) |0_0\rangle + ({\mathcal V}|1_1\rangle+|1_0\rangle ) |1_0\rangle ) \otimes  ({\mathcal U}|0_1\rangle+|0_0\rangle)]
\end{align}
up to normalization.  Here, we have defined ${\mathcal V} = \sum_{b,\beta,s = 0,1}\left|b_s\right> \left<b \right| V \left| \beta \right> \left<{\beta}_s\right|$ that applies $V$ while keeping the stage fixed.  If a circuit were to include more one-qubit gates, for each gate we would include another factor similar to the form $(({\mathcal V}|0_1\rangle+|0_0\rangle) |0_0\rangle + ({\mathcal V}|1_1\rangle+|1_0\rangle) |1_0\rangle )$ and perform an additional projection using $\hat{A}$.

The final example, shown in Fig. \ref{fig:Hamiltonians}(d), has a ground state of the form 
\begin{align}
(I^{(2)} \otimes \hat{A} \otimes I^{(2)} \otimes \hat{A}) & [ (|0_0\rangle |0_0\rangle + |1_0\rangle |1_0\rangle ) \otimes  |0_0\rangle \otimes (|0_0\rangle |0_0\rangle + |1_0\rangle |1_0\rangle ) \otimes  |0_0\rangle \nonumber \\
 + \sum_{b^\prime,B^\prime} & \langle b^\prime| \langle B^\prime |W | 0\rangle| 0\rangle (|0_0\rangle |0_0\rangle + |1_0\rangle |1_0\rangle ) \otimes  |b^\prime_1\rangle \otimes (|0_0\rangle |0_0\rangle + |1_0\rangle |1_0\rangle ) \otimes  |B^\prime_1\rangle].
\end{align}
The first term in brackets has both qubits in their initialized states $|0_0\rangle$ and $|0_0\rangle$.  The second term in brackets has both qubits emerging from $W$ in the states $ |b^\prime_1\rangle$ and  $|B^\prime_1\rangle$, with the transition matrix elements given by $\langle b^\prime| \langle B^\prime |W | 0\rangle| 0\rangle$.  The projection operators $\hat{A}$ take care of the teleportation circuits that act after $W$.

In these examples, we see the structure that characterizes $|\Psi(\theta)\rangle$ for any circuit.  The unitary gates are included in the entangled pairs, and the projections are performed using $\hat{A}$.

\twocolumngrid